\documentclass[%
 reprint,
superscriptaddress,
 amsmath,amssymb,
 aps,
]{revtex4-2}

\usepackage{graphicx}
\usepackage{dcolumn}
\usepackage{bm}

\usepackage{upgreek}

\begin{document}

\title{Theory of space-time supermodes in planar multimode waveguides}

\author{Abbas Shiri}
\affiliation{CREOL, The College of Optics \& Photonics, University of Central Florida, Orlando, FL 32816, USA}
\affiliation{Department of Electrical and Computer Engineering, University of Central Florida, Orlando, FL 32816, USA}

\author{Kenneth L. Schepler}
\affiliation{CREOL, The College of Optics \& Photonics, University of Central Florida, Orlando, FL 32816, USA}

\author{Ayman F. Abouraddy}
\affiliation{CREOL, The College of Optics \& Photonics, University of Central Florida, Orlando, FL 32816, USA}
\affiliation{Department of Electrical and Computer Engineering, University of Central Florida, Orlando, FL 32816, USA}
\email{*raddy@creol.ucf.edu}




\begin{abstract}
When an optical pulse is focused into a multimode waveguide or fiber, the energy is divided among the available guided modes. Consequently, the initially localized intensity spreads transversely, the spatial profile undergoes rapid variations with axial propagation, and the pulse disperses temporally. Space-time (ST) supermodes are pulsed guided field configurations that propagate invariantly in multimode waveguides by assigning each mode to a prescribed wavelength. ST supermodes can be thus viewed as spectrally discrete, guided-wave counterpart of the recently demonstrated propagation-invariant ST wave packets in free space. The group velocity of an ST supermode is tunable independently -- in principle -- of the waveguide structure, group-velocity dispersion is eliminated or dramatically curtailed, and the time-averaged intensity profile is axially invariant along the waveguide in absence of mode-coupling. We establish here a theoretical framework for studying ST supermodes in planar waveguides. Modal engineering allows sculpting this axially invariant transverse intensity profile from an on-axis peak or dip (dark beam), to a multi-peak or flat distribution. Moreover, ST supermodes can be synthesized using spectrally incoherent light, thus paving the way to potential applications in optical beam delivery for lighting applications.
\end{abstract}

\maketitle

\section{Introduction}

Space-time wave packets (STWPs) are pulsed beams endowed with a precise spatio-temporal structure \cite{Yessenov22AOP} that results in rigid transport (diffraction-free and dispersion-free propagation) in linear media \cite{Reivelt2003arxiv,Turunen2010PO,Figueroa2014Book,Hall23OL1km}. By assigning each spatial frequency underpinning the transverse spatial profile to one temporal frequency underpinning the pulse profile \cite{Donnelly1993ProcRSLA,Saari2004PRE,Kondakci2016OE,Parker2016OE,Porras2017OL,Kondakci2017NP}, a host of unique attributes can be attained in the synthesized STWP, such as self healing \cite{Kondakci2018OL}, tunable group velocity \cite{Salo2001JOA,Valtna2007OC,Efremidis2017OL,Wong2017ACSP2,Kondakci2019NC}, and anomalous refraction \cite{Bhaduri2020NP}. Historically, the first discovered STWP was Brittingham's focus-wave mode (FWM) \cite{Brittingham1983JAP}, which is an example of `sideband' STWPs, so-called because low spatial frequencies are excluded from the spectrum on physical grounds \cite{Yessenov2019PRA}. Subsequently, X-waves were identified and realized \cite{Lu1992IEEETUFFC-Xwaves,Saari1997LP,Saari1997PRL}. However, both sideband STWPs and X-waves are difficult to synthesize \cite{Saari1997LP,Reivelt2000JOSAA,Reivelt2002PRE,Turunen2010PO}, and any significant tuning of their attributes can be achieved only in the non-paraxial regime while utilizing ultrabroad bandwidths \cite{Yessenov22AOP}. More recently, the newly identified family of `baseband' STWPs has been pursued, whose characteristics can be tuned over unprecedented scales while using narrow bandwidths and remaining in the paraxial regime \cite{Kondakci2017NP,Kondakci2019NC,Yessenov2019OE}.

Until very recently, all experimental investigations of STWPs have been confined to freely propagating fields \cite{Turunen2010PO,Figueroa2014Book,Yessenov22AOP}, although a few theoretical studies examined the propagation of X-waves in multimode cylindrical waveguides  \cite{ZamboniRached2001PRE,ZamboniRached2002PRE,ZamboniRached2003PRE}, in addition to an early theoretical investigation of FWMs in a single-mode fiber \cite{Vengsarkar1992JOSAA}. More recently, the impact of orbital angular momentum (OAM) on FWMs in a single-mode fiber has been studied \cite{Ruano21JO}. In light of the above-mentioned difficulties in synthesizing X-waves and sideband STWPs (such as FWMs) \cite{Yessenov22AOP}, it is unlikely that these proposals can be put to test. However, recent studies have been directed at confining the more useful \textit{baseband} STWPs in waveguides. For example, hybrid guided space-time modes in single-mode \cite{Shiri2020NC_Hybrid} or few-mode planar waveguides \cite{Shiri2022ACSP} have been demonstrated in which the STWP structure is introduced along the unconfined dimension. In the context of highly multimoded waveguides, the propagation and excitation of STWPs via high-energy pulses in a step-index noninear multimode fiber or waveguide has been investigated theoretically \cite{Kibler21PRL,Bejot21ACSP,Bejot22ACSP}, and has very recently been realized experimentally \cite{Stefanska22arxiv}. Additionally, STWPs can be generated in a multimode planar, graded-index waveguide by endowing a segment of the waveguide with an appropriately designed time-varying refractive index \cite{Guo21PRR}.

These rapid developments in guided STWPs over the past two years raise a natural question: are STWPs propagation invariant in multimode waveguides just as they are in bulk media? Because guided modes have different propagation constants and group velocities, the spatial profile of a conventional pulsed multimode field changes erratically with propagation, even in absence of mode-coupling. In contrast, a recent proof-of-principle experiment \cite{Shiri2022STsupermodes} has indicated that an appropriately prepared, quasi-discrete-spectrum STWP in free space can couple to a multimode step-index planar waveguide and maintain its spatial profile along the waveguide length.

Here we develop the theory of STWPs in planar multimode optical waveguides in absence of mode coupling. We denote such guided optical fields `ST supermodes'. Ideal ST supermodes are formed of a superposition of monochromatic waveguide modes each at a prescribed frequency, which is selected to guarantee that the ST supermode as a whole propagates invariantly at a tunable group velocity. However, ideal ST supermodes have a strictly discretized spectrum and thus require infinite energy for their realization. We relax here the monochromaticity constraint to introduce realistic finite-energy ST supermodes, and also examine spectrally continuous ST supermodes. In all three cases (ideal, finite-energy, and spectrally continuous), the supermode group velocity is tunable, the impact of dispersion is reduced with respect to a conventional pulsed multimode field comprising the same modes, and the time-averaged intensity profile remains axially invariant as long as the spectra associated with the different modes do not overlap. Furthermore, the axially invariant intensity profile can be sculpted by varying the modal contributions, thereby producing an on-axis peak or dip, a multi-peak distribution, or a flat intensity profile. Moreover, we show that the axial invariance of this intensity profile extends to spectrally incoherent fields, which suggests potential applications in lighting. These results pave the way towards investigating ST supermodes in optical fibers and conventional waveguides in which light is confined along both transverse dimensions.

\section{overview of freely propagating space-time wave packets}\label{Section:ST_freespace}

We start by briefly reviewing STWPs in isotropic, non-dispersive media, and restrict the transverse spatial profile to one dimension $x$, while holding the field uniform along $y$, in anticipation of the treatment for ST supermodes in a planar waveguide. The field for a generic scalar pulsed beam (or wave packet) in a medium of refractive index $n$ is $E(x,z;t)\!=\!e^{i(nk_{\mathrm{o}}z-\omega_{\mathrm{o}}t)}\psi(x,z;t)$, where $\omega_{\mathrm{o}}$ is the carrier frequency, $k_{\mathrm{o}}\!=\!\tfrac{\omega_{\mathrm{o}}}{c}$ the associated free-space wave number, $c$ the speed of light in vacuum, and the angular spectrum for the envelope $\psi(x,z;t)$ has the form:
\begin{equation}
\psi(x,z;t)=\!\int\!\!\!\int dk_{x}d\Omega\,\widetilde{\psi}(k_{x},\Omega)e^{ik_{x}x}e^{i(k_{z}-nk_{\mathrm{o}})z}e^{-i\Omega t};
\end{equation}
here $\Omega\!=\!\omega-\omega_{\mathrm{o}}$, and the spatio-temporal spectrum $\widetilde{\psi}(k_{x},\Omega)$ is the 2D Fourier transform of $\psi(x,0;t)$, whose support on the surface of the light-cone $k_{x}^{2}+k_{z}^{2}\!=\!(n\tfrac{\omega}{c})^{2}$ is thus a 2D domain.

\begin{figure}[t!]
    \centering
    \includegraphics[width=8.4cm]{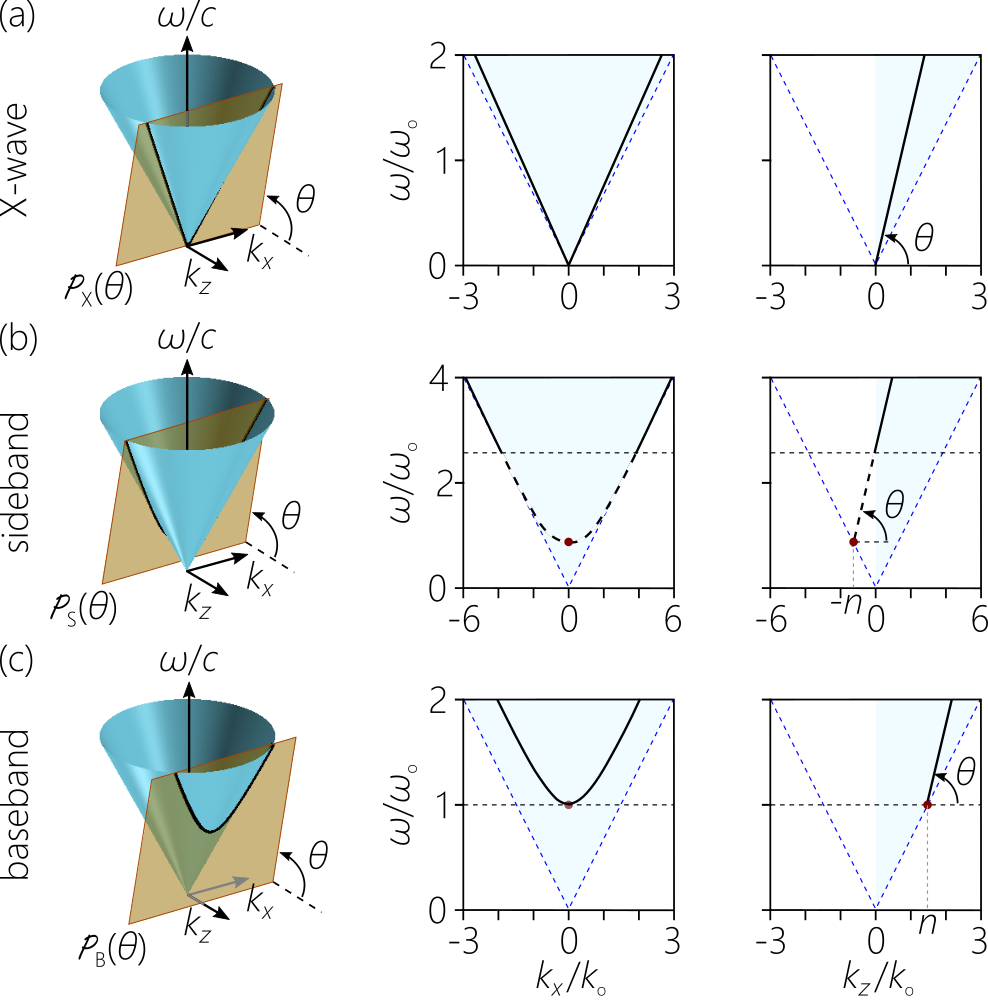}
    \caption{Spectral representation of freely propagating STWPs in a non-dispersive medium of refractive index $n$ ($n\!=\!1.5$ throughout). (a) X-waves. We depict the spectral support for an X-wave at the intersection of the light-cone $k_{x}^{2}+k_{z}^{2}\!=\!(n\tfrac{\omega}{c})^{2}$ with the spectral plane $\mathcal{P}_{X}(\theta)$, along with its spectral projections onto the $(k_{z},\tfrac{\omega}{c})$-plane, which is a straight line making an angle $\theta$ with the $k_{z}$-axis, and the projection onto the $(k_{x},\tfrac{\omega}{c})$-plane, which is a conic section \cite{Kondakci2017NP}. The shaded regions represent non-evanescent plane waves consistent with causal excitation and propagation \cite{Yessenov2019PRA,Yessenov22AOP}. Here $\omega_{\mathrm{o}}$ is an arbitrary normalization frequency. (b) Same as (a) for sideband STWPs. The dashed parts of the spectral projections are excluded because they correspond to negative values of $k_{z}$. (c) Same as (a) for baseband STWPs.}
    \label{Fig:LightCones}
\end{figure}

In contrast, the spectral support for STWPs [Fig.~\ref{Fig:LightCones}] is a 1D curve on the light-cone surface \cite{Donnelly1993ProcRSLA,Yessenov2019PRA}. The propagation invariance of STWPs requires that these spectral curves are the intersection of the light-cone with a spectral plane $\mathcal{P}(\theta)$ that is parallel to the $k_{x}$-axis. Consequently, the spectral projection for any STWP onto the $(k_{z},\tfrac{\omega}{c})$-plane is a straight line, thus signifying dispersionless propagation.

There are three basic types of STWPs: X-waves, sideband STWPs, and baseband STWPs \cite{Yessenov2019PRA}, which are distinguished by the form of $\mathcal{P}(\theta)$. For X-waves, the plane $\mathcal{P}_{\mathrm{X}}(\theta)$ passes through the origin $\omega\!=\!k_{z}c\tan{\theta}$, with $\tan^{-1}{(\tfrac{1}{n})}\!<\!\theta\!<\!\tfrac{\pi}{2}$ [Fig.~\ref{Fig:LightCones}(a)],  where $\theta$ (the spectral tilt angle) is the angle the plane makes with the $k_{z}$-axis. The field for the X-wave is:
\begin{equation}
E_{\mathrm{X}}(x,z;t)=\int\!d\Omega\,\widetilde{\psi}(\omega)e^{ik_{x}x}e^{-i\omega(t-z/\widetilde{v})}\!=\!E_{\mathrm{X}}(x,0;t-z/\widetilde{v}),
\end{equation}
where $\widetilde{v}\!=\!c\tan{\theta}$ is both the group and phase velocity of the rigidly transported X-wave, which is always superluminal ($\widetilde{v}\!>\!c/n$) \cite{Figueroa2014Book}. In the luminal limit $\widetilde{v}\!=\!c/n$, the X-wave degenerates into a plane-wave pulse \cite{Yessenov2019PRA}.

For \textit{sideband} STWPs, the spectral plane $\mathcal{P}_{\mathrm{S}}(\theta)$ is $\Omega\!=\!(k_{z}+nk_{\mathrm{o}})\widetilde{v}$, with group velocity $\widetilde{v}\!=\!c\tan{\theta}$ and $\tan^{-1}{(\tfrac{1}{n})}\!\leq\!\theta\!<\!\tfrac{\pi}{2}$, thus encompassing the luminal ($\widetilde{v}\!=\!c/n$) and superluminal $\widetilde{v}$ ($\widetilde{v}\!>\!c/n$) regimes [Fig.~\ref{Fig:LightCones}(b)]. The field is given by $E_{\mathrm{S}}(x,z;t)\!=\!e^{-i(nk_{\mathrm{o}}z+\omega_{\mathrm{o}}t)}\psi_{\mathrm{S}}(x,z;t)$, so that the phase velocity is $-c/n$. It is clear from Fig.~\ref{Fig:LightCones}(b) that the spatial frequencies $k_{x}$ in the vicinity of $k_{x}\!=\!0$ correspond to negative values of $k_{z}$, which is incompatible with causal excitation and propagation \cite{Heyman1987JOSAA,Turunen2010PO,Yessenov22AOP}. Consequently, the span of low spatial frequencies is excluded from sideband STWPs on physical grounds. Both X-waves and sideband STWPs require large numerical apertures and large bandwidths for their realization if $\widetilde{v}$ is to depart from $c/n$. Hence these two types of STWPs are difficult to produce experimentally \cite{Yessenov2019PRA,Yessenov22AOP}, and have found no optical applications in the $\sim\!3-4$ decades since their discovery \cite{Turunen2010PO}.

Finally, for \textit{baseband} STWPs the plane $\mathcal{P}_{\mathrm{B}}(\theta)$ is $\Omega\!=\!(k_{z}-nk_{\mathrm{o}})\widetilde{v}$, with $\widetilde{v}\!=\!c\tan{\theta}$ and $0\!<\!\theta\!<\!\pi$ [Fig.~\ref{Fig:LightCones}(c)]. Thus the group velocity of baseband STWPs can uniquely take on subluminal, superluminal, or negative values \cite{Kondakci2019NC}. The field is given by $E_{\mathrm{B}}(x,z;t)\!=\!e^{i(nk_{\mathrm{o}}z-\omega_{\mathrm{o}}t)}\psi_{\mathrm{B}}(x,z;t)$, so that the phase velocity is $c/n$. The angular spectrum for the envelopes of sideband and baseband STWPs have the same form:
\begin{equation}\label{Eq:STWPEnvelope}
\psi(x,z;t)=\int\!d\Omega\,\widetilde{\psi}(\Omega)e^{ik_{x}x}e^{-i\Omega(t-z/\widetilde{v})}\!=\!\psi(x,0;t-z/\widetilde{v}).
\end{equation}
Both types of STWPs thus propagate invariantly at a group velocity $\widetilde{v}\!=\!c\tan{\theta}$. However, there are two crucial distinctions. First, sideband STWPs can be only luminal (FWMs) or superluminal, whereas the group velocity of  baseband STWPs can take on -- in principle -- arbitrary values. Second, the spatial spectrum of baseband STWPs is in the vicinity of $k_{x}\!=\!0$ [Fig.~\ref{Fig:LightCones}(c)], whereas that for sideband STWPs [Fig.~\ref{Fig:LightCones}(b)] is displaced away from $k_{x}\!=\!0$ \cite{Yessenov2019PRA,Yessenov22AOP}.

\section{ST supermodes in planar waveguides}\label{Section:ST_waveguides}

\subsection{Spectral representation of planar-waveguide modes}

\begin{figure*}[t!]
    \centering
    \includegraphics[width=16cm]{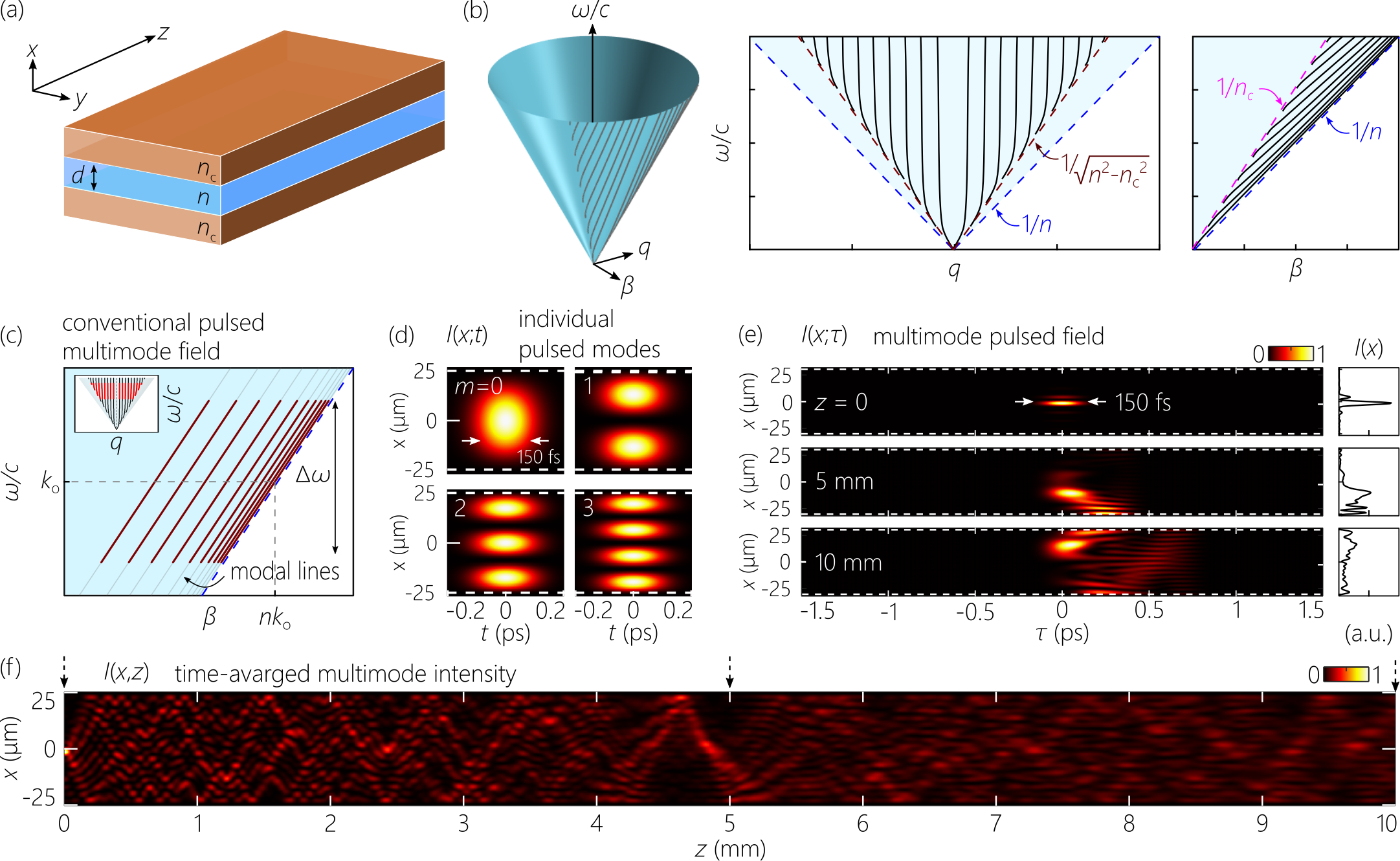}
    \caption{(a) Schematic of a planar step-index waveguide with core and cladding refractive indices $n$ and $n_{\mathrm{c}}$, respectively. (b) The spectral support for the dispersion relationships $\beta_{m}(\omega)$ associated with the first 10 TE waveguide modes are 1D trajectories on the light-cone surface $\beta^{2}+q^{2}\!=\!(n\tfrac{\omega}{c})^{2}$. We also plot their spectral projections (solid black curves) onto the $(q,\tfrac{\omega}{c})$ and $(\beta,\tfrac{\omega}{c})$ planes. The blue-shaded regions (similarly to Fig.~\ref{Fig:LightCones}) correspond to propagating field components in the bulk. (c) A pulsed localized optical field at the input of a multimode waveguide couples to multiple modes over a bandwidth $\Delta\omega$. The inset depicts the corresponding spectral projection onto the $(q,\tfrac{\omega}{c})$-plane. (d) The spatio-temporal intensity profiles $I(x;t)$ for individual pulsed modes at $z\!=\!0$; $\lambda_{\mathrm{o}}\!=\!1.55$~$\upmu$m, $\Delta T\!=\!150$~fs, and $\Delta\lambda\!\approx\!23$~nm. (e) The spatio-temporal intensity $I(x,z;\tau)$, with $\tau\!=\!t-nz/c$, for the pulsed multimode field comprising an equal-weighted superposition of the first 16 modes in a frame moving at $v\!=\!c/n$ at $z\!=\!0$, 5, and 10~mm. (f) The time-averaged intensity $I(x,z)$ of the field in (e). The axial planes selected in (e) are identified by dashed arrows.}
    \label{Fig:MultimodeWaveguide}
\end{figure*}

Consider the symmetric step-index planar dielectric waveguide structure depicted in Fig.~\ref{Fig:MultimodeWaveguide}(a), with core refractive index $n$ and cladding $n_{\mathrm{c}}\!<\!n$. The index-guided field $E(x,z;t)$ propagates along $z$, is confined along $x$, and is uniform along $y$. We refer to the axial and transverse wave numbers as $\beta$ and $q$, respectively, so that the light-cone surface for the core medium is $\beta^{2}+q^{2}\!=\!(n\tfrac{\omega}{c})^{2}$. For the $m^{\mathrm{th}}$-order mode ($m\!=\!0,1,2,\cdots$), the axial and transverse wave numbers in the core are $\beta_{m}$ and $q_{m}$, respectively. The spectral projection onto the $(\beta,\tfrac{\omega}{c})$-plane for any modal dispersion relationship $\beta_{m}(\omega)$ is thus a 1D curve [Fig.~\ref{Fig:MultimodeWaveguide}(b)]. However, in contrast to STWPs, this spectral projection is \textit{not} a straight line, so that guided modes are always dispersive. For comparison, we consider in the Appendix the analytically tractable case of a planar waveguide formed of two perfect mirrors, in which case closed-form expressions can be derived for the quantities we compute here.

\subsection{Conventional pulsed multimode guided fields}

In a multimode waveguide, the field is a superposition of modes $E(x,z;t)\!=\!\sum_{m}A_{m}E_{m}(x,z;t)$, where $A_{m}$ is the modal amplitude, normalized such that $\sum_{m}|A_{m}|^{2}\!=\!1$, and $E_{m}(x,z;t)$ is the field for the $m^{\mathrm{th}}$ mode, normalized such that $\iint\!dxdt\,|E(x,z;t)|^{2}\!=\!1$ in absence of absorption or radiation leakage. At a fixed frequency $\omega_{\mathrm{o}}$, the monochromatic modal fields are $E_{m}(x,z;t)\!=\!u_{m}(x;\omega_{\mathrm{o}})e^{i\{\beta_{m}(\omega_{\mathrm{o}})z-\omega_{\mathrm{o}}t\}}$, where $u_{m}(x;\omega_{\mathrm{o}})$ is the transverse spatial mode profile. We consider throughout sufficiently small bandwidths $\Delta\omega$ so that one may ignore the frequency dependence of the modal profiles, $u_{m}(x;\omega)\!\approx\!u_{m}(x)$, which are orthonormal $\int\!dx\,u_{m}(x)u_{\ell}^{*}(x)\!=\!\delta_{m\ell}$. The spectral support for this monochromatic multimode field is a collection of points on the light-cone surface at the intersection of the horizontal iso-frequency plane $\omega\!=\!\omega_{\mathrm{o}}$ with modal dispersion curves $\beta_{m}(\omega)$ shown in Fig.~\ref{Fig:MultimodeWaveguide}(b). The intensity of this monochromatic multimode field is:
\begin{equation}
I(x,z)=|E(x,z;t)|^{2}=|\sum_{m}A_{m}u_{m}(x)e^{i\beta_{m}(\omega_{\mathrm{o}})z}|^{2}.
\end{equation}
Starting with a localized field at $z\!=\!0$, the axially varying relative phase factor $e^{i\beta_{m}(\omega_{\mathrm{o}})z}$ produces erratic changes in the transverse intensity profile along $z$.

If instead a \textit{pulse} (width $\Delta T$ and bandwidth $\Delta\omega\!\sim\!\tfrac{1}{\Delta T}$ centered at $\omega_{\mathrm{o}}$) is spatially localized at $z\!=\!0$, the pulse energy once coupled to the waveguide is typically distributed among a multiplicity of modes, and we assume, for simplicity, that the excited modes share the same bandwidth $\Delta\omega$ [Fig.~\ref{Fig:MultimodeWaveguide}(c)]. The individual pulsed modes can be expressed as:
\begin{equation}\label{Eq:ModeConventionalPulsed}
E_{m}(x,z;t)=e^{i\{\beta_{m}(\omega_{\mathrm{o}})z-\omega_{\mathrm{o}}t\}}u_{m}(x)\psi_{m}(z;t),
\end{equation}
where $\psi_{m}(z;t)$ is an axial modal envelope. By expanding $\beta_{m}(\omega)$ into a Taylor series around $\omega_{\mathrm{o}}$, $\beta_{m}(\omega)\!=\!\beta_{m}(\omega_{\mathrm{o}}+\Omega)=\beta_{m}(\omega_{\mathrm{o}})+\frac{\Omega}{\widetilde{v}_{m}}+\frac{1}{2}\eta_{m}\Omega^{2}+\cdots$, we can write $\psi_{m}(z;t)$ as follows:
\begin{equation}\label{Eq:GeneralModalAxialEnvelope}
\psi_{m}(z;t)=\int_{\Delta\omega}\!\!d\Omega\,\widetilde{\psi}(\Omega)\,e^{-i\Omega(t-z/\widetilde{v}_{m})}e^{i\eta_{m}\Omega^{2}z/2};
\end{equation}
here $\Omega\!=\!\omega-\omega_{\mathrm{o}}$, $\widetilde{\psi}(\Omega)$ is the complex spectrum of the input pulse, the integration domain is its full bandwidth $\Delta\omega$, $\widetilde{v}_{m}\!=\!1\big/\tfrac{d\beta_{m}}{d\omega}\big|_{\omega_{\mathrm{o}}}$ is the group velocity of the $m^{\mathrm{th}}$-mode, its phase velocity is $v_{m}\!=\!\tfrac{\omega_{\mathrm{o}}}{\beta_{m}(\omega_{\mathrm{o}})}$, and $\eta_{m}\!=\!\tfrac{d^{2}\beta_{m}}{d\omega^{2}}\big|_{\omega_{\mathrm{o}}}$ is the group-velocity dispersion (GVD) coefficient \cite{Saleh2007Book}. We normalize the spectrum so that $\int\!dt|\psi_{m}(z;t)|^{2}\!=\!1$. In absence of GVD ($\eta_{m}\!=\!0$), the envelope has a temporal width of $\Delta T$ traveling rigidly at a group velocity $\widetilde{v}_{m}$. In presence of GVD, the envelope disperses temporally with a chirp parameter $a_{m}\!=\!\eta_{m}(\Delta\omega)^{2}z$ \cite{Saleh2007Book}. We plot in Fig.~\ref{Fig:MultimodeWaveguide}(d) examples of the spatio-temporal intensity profiles for individual pulsed modes at the waveguide entrance $I(x;t)\!=\!|E_{m}(x,z\!=\!0;t)|^{2}$ for $m\!=\!0,1,2,3$, which are \textit{separable} with respect to $x$ and $t$. Although the spatio-temporal profile $I_{m}(x,z;t)\!=\!|E_{m}(x,z;t)|^{2}$ for a single pulsed mode changes along $z$ due to dispersion, the time-averaged intensity $I_{m}(x,z)\!=\!\int\!dt\,I_{m}(x,z;t)\!=\!|u_{m}(x)|^{2}$ is independent of $z$.

The initially localized separable pulsed multimode field at $z\!=\!0$ is a superposition of such pulsed modes. The field subsequently undergoes complex propagation dynamics, in part due to the relative phase factors $e^{i\beta_{m}(\omega_{\mathrm{o}})z}$, similarly to the monochromatic multimode fields described above. Additionally, the pulsed multimode field comprises modal envelopes traveling at different group velocities $\widetilde{v}_{m}$ (modal GVD), each separately undergoing dispersive spreading with a chirp parameter $a_{m}$ (waveguide GVD), and potentially chromatic GVD (if $n$ is wavelength-dependent). An example is shown in Fig.~\ref{Fig:MultimodeWaveguide}(e) of an equal-weight superposition of the first 16~modes. At $z\!=\!0$, the input field is localized along $x$ (beam width $\Delta x\!=\!5.6$~$\upmu$m) and along $t$ (pulse width $\Delta T\!=\!150$~fs). Subsequently, the pulsed multimode field spreads temporally and spatially [Fig.~\ref{Fig:MultimodeWaveguide}(e)].

The time-averaged intensity for this pulsed multimode field, $I(x,z)\!=\!\int\!dt\,|E(x,z;t)|^{2}$, is:
\begin{equation}\label{Eq:ConvMultimodePulsedIntensity}
I(x,z)=\sum_{m,\ell}A_{m}A_{\ell}^{*}u_{m}(x)u_{\ell}^{*}(x)\psi_{m\ell}(z),
\end{equation}
where the mutual envelope $\psi_{m\ell}(z)$ is:
\begin{equation}
\psi_{m\ell}(z)=\int\!d\Omega\,|\widetilde{\psi}(\Omega)|^{2}\,e^{i\Omega(\widetilde{n}_{m}-\widetilde{n}_{\ell})z/c}e^{i(\eta_{m}-\eta_{\ell})\Omega^{2}z/2};
\end{equation}
here $\widetilde{n}_{m}\!=\!c/\widetilde{v}_{m}$ is the modal group index. The intensity $I(x,z)$ is initially localized in the vicinity of $x\!=\!0$ at $z\!=\!0$ but expands rapidly to the waveguide interfaces, and the transverse profile subsequently varies erratically along $z$ [Fig.~\ref{Fig:MultimodeWaveguide}(f)]. This is a generic feature of all multimode waveguides of sufficiently large size. We proceed to show that ST supermodes that combine these same modes across the same temporal bandwidth $\Delta\omega$ nevertheless retain their intensity profile invariantly along the waveguide.

\subsection{Ideal ST supermodes}

As shown in Fig.~\ref{Fig:LightCones}, the spectral support for a freely propagating STWP is the 1D curve at the intersection of the light-cone with a plane $\mathcal{P}(\theta)$. An STWP propagates invariantly without diffraction or dispersion in linear media \cite{Porras2003PREBessel-X,Longhi2004OL,Porras2004,Malaguti2008OL,Malaguti2009PRA,Hall2022LPR} at a group velocity $\widetilde{v}\!=\!c\tan{\theta}$, which can be changed -- in principle -- independently of the refractive index by tuning $\theta$. In a similar vein, the spectral support for a waveguide mode is a 1D curve on the light-cone surface. However, in contrast to STWPs, guided modes are dispersive, and their group velocity is determined by the waveguide structure. Therefore, pulsed single-mode and multimode guided fields do not propagate invariantly.

\begin{figure*}[t!]
\centering
\includegraphics[width=18cm]{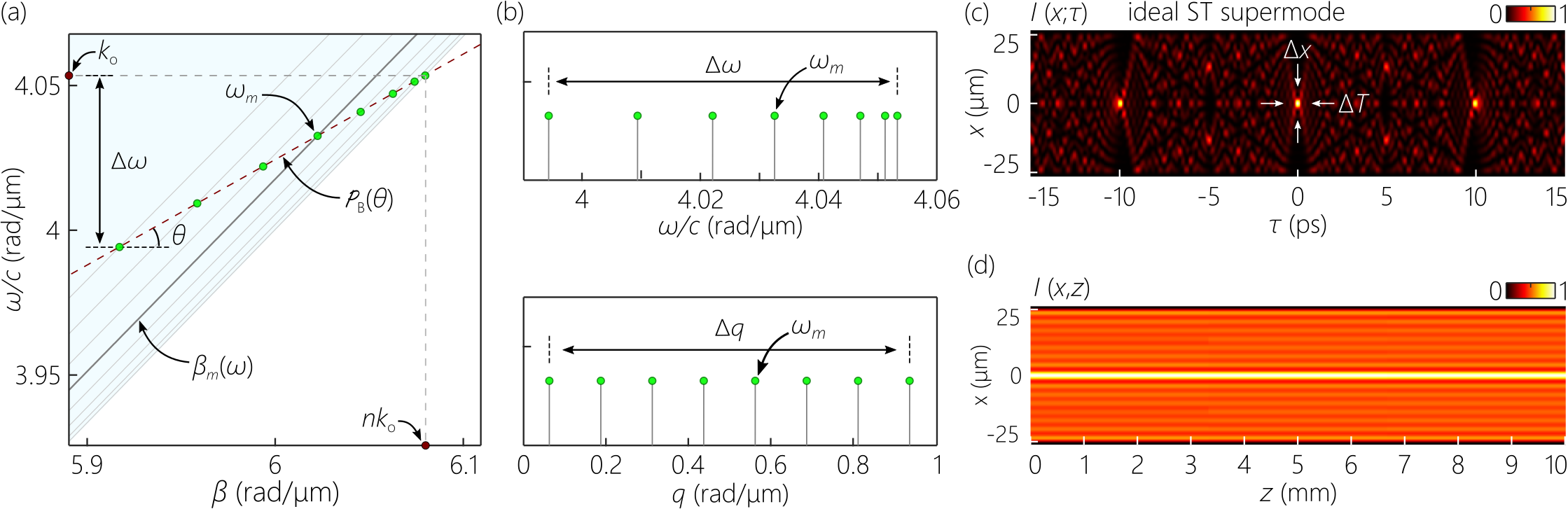}
\caption{Ideal ST supermodes. (a) Spectral projection onto the $(\beta,\tfrac{\omega}{c})$-plane of the modal dispersion curves $\beta_{m}(\omega)$ and the spectral plane $\mathcal{P}(\theta)$; here $\theta\!=\!20^{\circ}$. The spectral representation of an ideal supermode comprises the intersection points at the frequencies $\omega_{m}$ (represented by green dots). (b) Projected temporal and spatial spectra for an ideal ST supermode onto the $\omega$-axis and the $q$-axis. (c) Spatio-temporal intensity profile $I(x,z;\tau)$ for an ideal ST supermode at a fixed axial plane, obtained by an equal-weight superposition of the first 8 even-parity modes ($m\!=\!0,2,\cdots,14$), with $\Delta T\!=\!280$~fs, $\Delta\lambda\!\approx\!22.6$~nm, and $\Delta x\!=\!2.8$~$\upmu$m. (d) Time-averaged intensity $I(x,z)$ for the ideal ST supermode in (c).}
\label{Fig:IdealSTSupermode}
\end{figure*} 

An \textit{ideal ST supermode} is a pulsed multimode guided field that nevertheless propagates invariantly along the waveguide. This unique behavior is achieved by superposing a multiplicity of \textit{monochromatic} modes, each at a frequency $\omega_{m}$ selected such that the supermode envelope travels unchanged at a group velocity $\widetilde{v}$ that is distinct from those of any of the underlying modes, and which can be tuned -- in principle -- independently of the waveguide structure. The field for the ST supermode can thus be expressed as follows:
\begin{equation}
E_{\mathrm{ST}}(x,z;t)=\sum_{m}A_{m}E_{\mathrm{ST},m}(x,z;t),
\end{equation}
where the contribution from each mode is:
\begin{equation}
E_{\mathrm{ST},m}(x,z;t)=u_{m}(x)e^{i\{\beta_{m}(\omega_{m})z-\omega_{m}t\}},
\end{equation}
and the frequency $\omega_{m}$ is at the intersection of the modal dispersion curve $\beta_{m}(\omega)$ with a spectral plane $\mathcal{P}(\theta)$. Consequently, the spectrum of an idealized ST supermode is discretized. 

For ST supermodes based on baseband STWPs (referred to henceforth as baseband ST supermodes), the spectral plane $\mathcal{P}_{\mathrm{B}}(\theta)$ is $\omega\!=\!\omega_{\mathrm{o}}+(\beta-nk_{\mathrm{o}})\widetilde{v}$, where $\widetilde{v}\!=\!c\tan{\theta}$. Consequently, the axial wave number for any mode is $\beta_{m}(\omega_{m})\!=\!(\omega_{m}-\omega_{\mathrm{o}})/\widetilde{v}+nk_{\mathrm{o}}$, and the ST supermode field is $E_{\mathrm{ST}}(x,z;t)\!=\!e^{i(nk_{\mathrm{o}}z-\omega_{\mathrm{o}}t)}\psi_{\mathrm{ST}}(x,z;t)$, where the overall axially invariant envelope is given by:
\begin{equation}\label{Eq:RealisticSTSupermodeSuperposition}
\psi_{\mathrm{ST}}(x,z;t)=\sum_{m}u_{m}(x)e^{-i(\omega_{m}-\omega_{\mathrm{o}})(t-z/\widetilde{v})}=\psi_{\mathrm{ST}}(x,0;t-z/\widetilde{v}).
\end{equation}
The phase velocity is $v\!=\!c/n$ and the group velocity is $\widetilde{v}\!=\!c\tan{\theta}$. The spectral support for an ideal baseband ST supermode [Fig.~\ref{Fig:IdealSTSupermode}(a)] comprises a group of points (each representing a monochromatic mode) at the intersection of $\mathcal{P}_{\mathrm{B}}(\theta)$ with the modal dispersion curves $\beta_{m}(\omega)$ evaluated for the waveguide in Fig.~\ref{Fig:MultimodeWaveguide}(a).

Alternatively, an ST supermode can be based on sideband STWPs, in which case the frequencies $\omega_{m}$ of the contributing modes are determined by intersecting the modal dispersion curves with the spectral plane $\mathcal{P}_{\mathrm{S}}(\theta)$, $\omega\!=\!\omega_{\mathrm{o}}+(\beta+nk_{\mathrm{o}})\widetilde{v}$. The field for such a sideband ST supermode is $E_{\mathrm{ST}}(x,z;t)\!=\!e^{-i(nk_{\mathrm{o}}z+\omega_{\mathrm{o}}t)}\psi_{\mathrm{ST}}(x,z;t)$, so the phase velocity is $v\!=\!-c/n$, and the envelope $\psi_{\mathrm{ST}}(x,z;t)$ is given by Eq.~\ref{Eq:RealisticSTSupermodeSuperposition}, except that the values of $\omega_{m}$ and the range of values for $m$ will be different from those of a baseband ST supermode sharing the same values of $\theta$ and $\Delta\omega$. Finally, X-wave ST supermodes are formed by utilizing the plane $\mathcal{P}_{\mathrm{X}}(\theta)$ given by $\omega\!=\!\beta\widetilde{v}$, in which case:
\begin{equation}
E_{\mathrm{ST}}(x,z;t)=\sum_{m}u_{m}(x)e^{-i\omega_{m}(t-z/\widetilde{v})}=E_{\mathrm{ST}}(x,0;t-z/\widetilde{v}).
\end{equation}

Despite the differences between baseband, sideband, and X-wave ST supermodes, the spatio-temporal intensity is axially invariant $I(x,z;t)\!=\!|E_{\mathrm{ST}}(x,z;t)|^{2}\!=\!I(x,0;t-z/\widetilde{v})$, and their idealized spectra are discrete [Fig.~\ref{Fig:IdealSTSupermode}(b)] at frequencies $\omega_{m}$ extending over the bandwidth $\Delta\omega$. However, this ideal limit corresponds to infinite energy; that is, $\int\!dxdt\,|E_{\mathrm{ST}}(x,z;t)|^{2}$ is not bounded, and thus cannot be realized in practice. We will lift this restriction below in two ways: first by relaxing the monochromaticity constraint and allowing a finite but small bandwidth $\delta\omega$ (the spectral uncertainty) for each mode centered at the ideal modal frequencies $\omega_{m}$, and then by examining the opposite limit of ST supermodes with continuous spectra.

\begin{figure*}[t!]
\centering
\includegraphics[width=12.5cm]{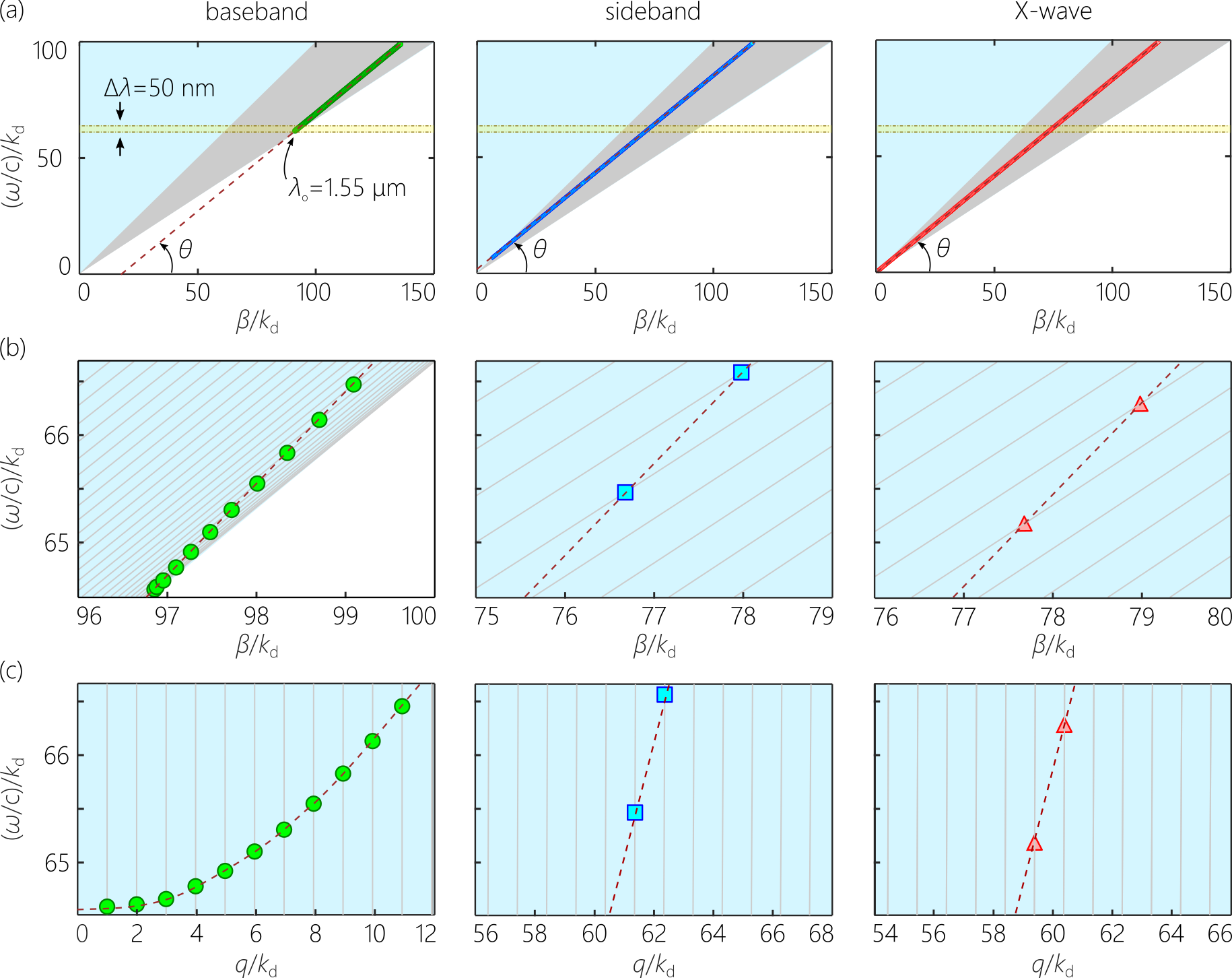}
\caption{Comparison between index-guided ST supermodes based on baseband STWPs, sideband STWPs, and X-waves in a planar multimode waveguide. In all cases $d\!=\!50$~$\upmu$m, $n\!=\!1.5$, and $n_{\mathrm{c}}\!=\!1$. (a) Intersection of the modal lines $\beta_{m}(\omega)$, represented by the grey-shaded zone, with a spectral plane (thin red dotted line at $\theta\!=\!40^{\circ}$) in the $(\beta,\frac{\omega}{c})$-plane. The horizontal and vertical axes are normalized with respect to $k_{d}\!=\!\tfrac{\pi}{d}$. (b) Same as (a) after zooming in onto a bandwidth $\Delta\lambda\!=\!50$~nm at a carrier wavelength of $\lambda_{\mathrm{o}}\!=\!1.55$~$\upmu$m, corresponding to the yellow horizontal zone between two dotted lines in (a). (c) Same as (b), but projected onto the $(q,\frac{\omega}{c})$-plane.}
\label{Fig:STSupermodeIntersections}
\end{figure*} 

\begin{figure*}
\centering
\includegraphics[width=18cm]{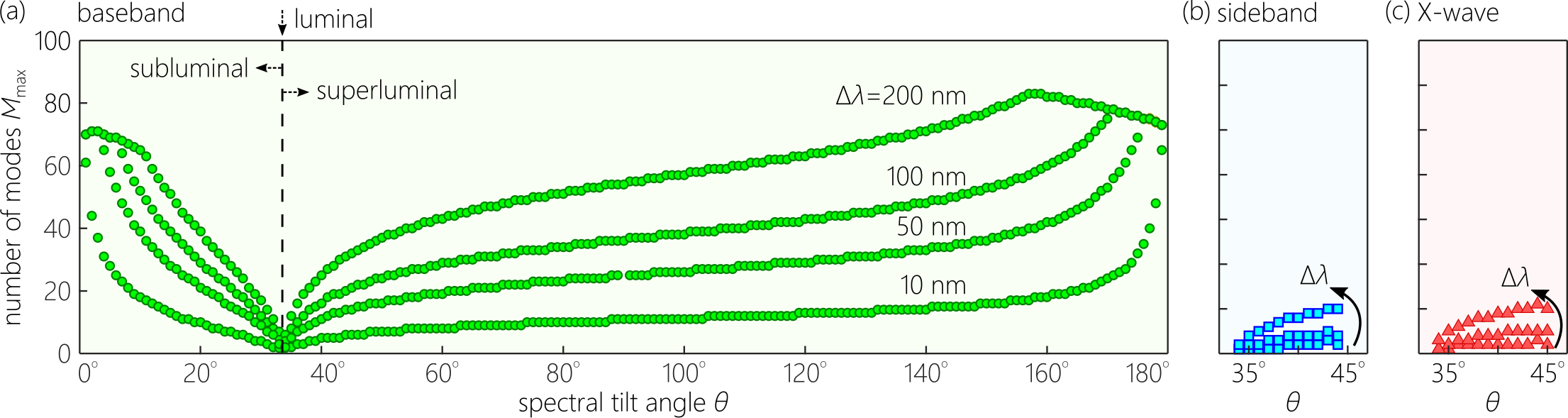}
\caption{(a) Maximum number of modes $M_{\mathrm{max}}$ accessible to the baseband ST supermodes in Fig.~\ref{Fig:STSupermodeIntersections} with the spectral tilt angle $\theta$, at bandwidths $\Delta\lambda\!=\!10$, 50, 100, and 200~nm; $\lambda_{\mathrm{o}}\!=\!1.55$~$\upmu$m, $n\!=\!1.5$, $n_{\mathrm{c}}\!=\!1$, and $d\!=\!50$~$\upmu$m. (b,c) Same as (a), but for (b) sideband and (c) X-wave ST supermodes. In these two cases, $\theta$ is limited by the core and cladding light-lines, from $\theta\!\approx\!34^{\circ}$ to $45^{\circ}$.}
\label{Fig:STSupermodeNumberOfModes}
\end{figure*}

We plot in Fig.~\ref{Fig:IdealSTSupermode}(c) the spatio-temporal intensity profile $I(x,z;t)\!=\!|E_{\mathrm{ST}}(x,z;t)|^{2}$ for an ideal baseband ST supermode at a fixed axial plane. This profile is independent of axial position $I(x,z;t)\!=\!I(x,0;t-z/\widetilde{v})$, and is formed of a central X-shaped spatio-temporal structure that resembles that of a freely propagating STWP. The temporal width $\Delta T$ of the peak of the X-shaped feature is determined by the temporal bandwidth ($\Delta T\!\sim\!\tfrac{1}{\Delta\omega}$), and its spatial width $\Delta x$ is determined by the spatial bandwidth ($\Delta x\!\sim\!\tfrac{1}{\Delta q}$). However, a field formed of the superposition of monochromatic modes will extend indefinitely in time around the central localized feature [Fig.~\ref{Fig:IdealSTSupermode}(c)]. A related phenomenon was recently observed for free STWPs after discretizing the spectrum in the context of space-time Talbot effects \cite{Yessenov2020PRLveiled,Hall2021APLSTTalbot}. Nevertheless, the intensity distribution is \textit{not} periodic in time because the frequencies $\omega_{m}$ are not equally spaced. We show below that relaxing the monochromaticity constraint reduces the overall temporal extent of the ST supermode around the central X-shaped feature.

The group velocity of the ST supermode $\widetilde{v}\!=\!c\tan{\theta}$ (group index $\widetilde{n}\!=\!\cot{\theta}$) is independent of the waveguide structure and is distinct from the group velocities of the contributing modes $\widetilde{v}_{m}$. This group velocity is the speed of the central feature in Fig.~\ref{Fig:IdealSTSupermode}(c), which can in principle be subluminal ($\widetilde{v}\!<\!c/n$), superluminal ($\widetilde{v}\!>\!c/n$), or even negative-valued. This group velocity can be tuned in the same waveguide by tuning $\theta$, which entails changing the frequencies $\omega_{m}$ assigned to the modes.

In the ideal limit considered here, all effects of GVD (whether modal, waveguide, or chromatic) have strictly speaking been eliminated. Because each contributing mode is monochromatic, waveguide GVD (and, likewise, chromatic GVD) is eliminated. Furthermore, the spectral projection for the ST supermode onto the $(\beta,\tfrac{\omega}{c})$-plane must lie along a straight line, thereby indicating a fixed group velocity $\widetilde{v}$, and thus elimination of modal GVD.

Finally, the time-averaged intensity $I(x,z)\!=\!\int\!dt\,|E_{\mathrm{ST}}(x,z;t)|^{2}$ can be readily shown to be:
\begin{equation}\label{Eq:TimeAveragedIntensityAxiallyInv}
I(x,z)=\sum_{m}|A_{m}|^{2}|u_{m}(x)|^{2}=I(x).
\end{equation}
In other words, the time-averaged intensity is fixed along the waveguide [Fig.~\ref{Fig:IdealSTSupermode}(d)], in contrast to conventional pulsed multimode fields having the same contributing modes and bandwidth $\Delta\omega$ [Eq.~\ref{Eq:ConvMultimodePulsedIntensity}, Fig.~\ref{Fig:MultimodeWaveguide}(f)]. By varying the modal weights $|A_{m}|^{2}$, one may indeed sculpt the axially invariant transverse intensity profile $I(x)$, which we explore further below.

\subsection{Comparison of baseband, sideband, and X-wave ST supermodes}

Any of the three types of STWPs (baseband, sideband, or X-waves) can in principle be utilized to construct an ST supermode. Indeed, early theoretical efforts considered making use of X-waves \cite{ZamboniRached2001PRE,ZamboniRached2002PRE,ZamboniRached2003PRE} and sideband FWMs \cite{Vengsarkar1992JOSAA} for this purpose. However, just as these two types of STWPs have not proven amenable to physical realization in free space at optical frequencies \cite{Yessenov22AOP}, they face similarly insurmountable technical hurdles for contributing to ST supermodes. We verify here that baseband STWPs provide the only practical path towards realizing ST supermodes, and that they are more versatile with respect to tuning the attributes of the ST supermode.

To elucidate the reason why baseband STWPs succeed where X-waves and sideband STWPs come up short, we consider the maximum number of waveguide modes $M_{\mathrm{max}}$ that can contribute to an ST supermode over a fixed temporal bandwidth $\Delta\omega$. Small $M_{\mathrm{max}}$ curtails the advantages conferred by ST supermodes, and the situation approaches that of attempting to couple to single mode or few modes in a highly multimoded waveguide. In Fig.~\ref{Fig:STSupermodeIntersections} we examine $M_{\mathrm{max}}$ for baseband, sideband, and X-wave ST supermodes, while holding the following key parameters fixed: $\theta\!=\!40^{\circ}$ and $\Delta\omega\!=\!2.15\tfrac{\pi c}{d}$, which corresponds to $\Delta\lambda\!=\!50$~nm at $\lambda_{\mathrm{o}}\!=\!1.55$~$\upmu$m and $d\!=\!50$~$\upmu$m. We select a superluminal group velocity $\widetilde{v}\!\approx\!0.84c\!>\!c/1.5$ because the superluminal regime is common to all three types of STWPs, whereas baseband STWPs uniquely span the subluminal \textit{and} negative-$\widetilde{v}$ regimes. Indeed, for sideband and X-wave ST supermodes we have $\tfrac{1}{n}\!<\!\tan{\theta}\!<\!\tfrac{1}{n_{\mathrm{c}}}$ ($33.7^{\circ}\!<\theta\!<\!45^{\circ}$) when $n\!=\!1.5$ and $n_{\mathrm{c}}\!=\!1$). This limitation is absent from baseband ST supermodes where the range for $\theta$ remains $0^{\circ}\!<\!\theta\!<\!180^{\circ}$ as for their freely propagating counterparts.

We identify in Fig.~\ref{Fig:STSupermodeIntersections}(a) the intersections of the spectral planes with the dispersion curves $\beta_{m}(\omega)$ projected onto the $(\beta,\tfrac{\omega}{c})$-plane. In Fig.~\ref{Fig:STSupermodeIntersections}(b) we zoom in onto the selected spectral window to evaluate $M_{\mathrm{max}}$. Although all three ST supermodes have the same $\theta$, the baseband ST supermode has a distinct advantage because its spectral support is in general closer to the light-line of the core medium. Consequently, the spectral plane intersects with the dispersion curves of a larger number of guided modes than in the case of the sideband or the X-wave ST supermode. Moreover, the lowest-order modes are associated with the baseband ST supermode (starting from $m\!=\!0$), whereas those associated with the other two types are high-order modes, as is clear in the spectral projections onto the $(q,\tfrac{\omega}{c})$-plane [Fig.~\ref{Fig:STSupermodeIntersections}(c)].

This conclusion is highlighted further in Fig.~\ref{Fig:STSupermodeNumberOfModes} where we plot $M_{\mathrm{max}}$ with $\theta$ at different bandwidths for all three types of ST supermodes. In all cases, $M_{\mathrm{max}}$ increases as $\theta$ deviates away from the luminal condition ($\widetilde{n}\!=\!n$). Moreover, $M_{\mathrm{max}}$ increases with bandwidth at fixed $\theta$. For any selection of parameters, $M_{\mathrm{max}}$ is always significantly larger for the baseband case when compared with the sideband or X-wave cases. Additionally, the baseband ST supermode enables spanning a much wider range of values for $\theta$. Therefore, baseband ST supermodes can have access to a larger number of modes $M_{\mathrm{max}}$ for a fixed $\Delta\omega$ and $\theta$, they can make use of the low-order modes, and thus require a smaller numerical aperture than that for sideband or X-wave ST supermodes. In light of these findings, we proceed to examine solely baseband ST supermodes, which are likely to be the only ones realizable in practice \cite{Shiri2022STsupermodes}.

\section{Finite-energy ST supermodes}

\begin{figure*}[t!]
\centering
\includegraphics[width=18cm]{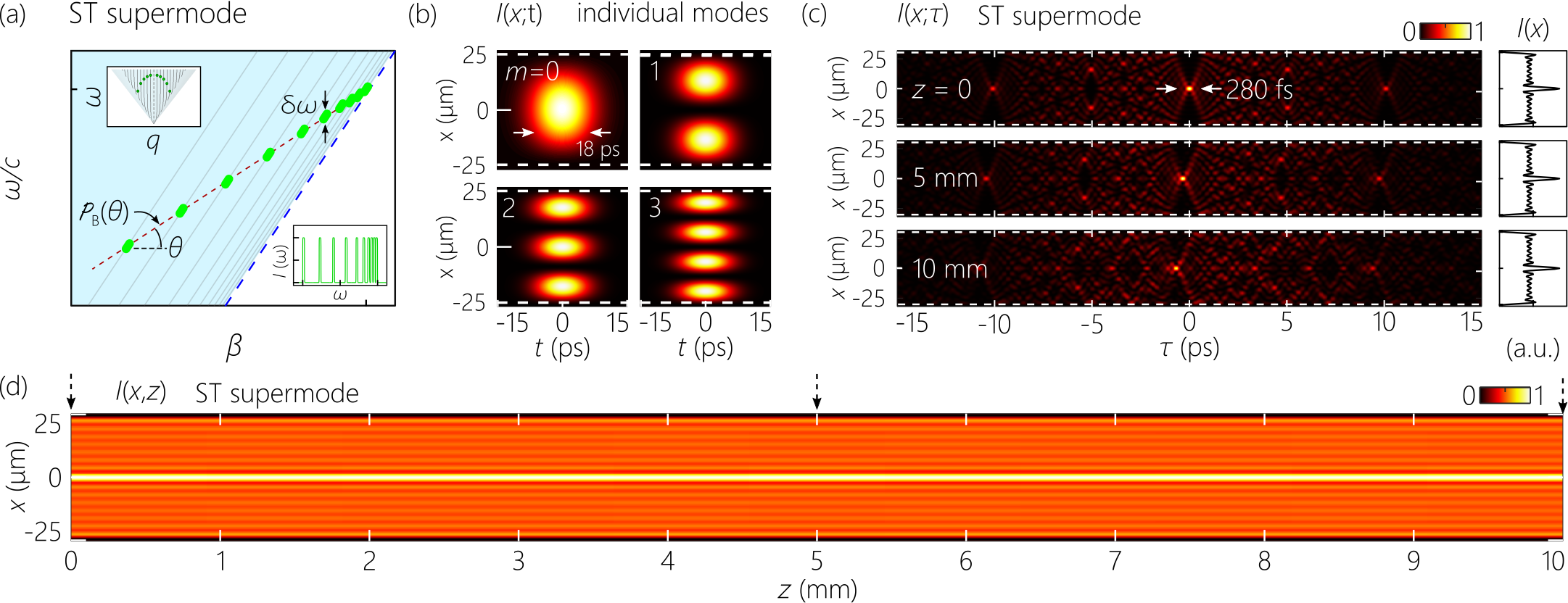}
\caption{(a) The spectral support for a finite-energy baseband ST supermode (the small green domains) at the intersection of the modal dispersion curves $\beta_{m}(\omega)$ for the waveguide in Fig.~\ref{Fig:MultimodeWaveguide}(a) with a tilted spectral plane $\omega\!=\!\omega_{\mathrm{o}}+(\beta-nk_{\mathrm{o}})\widetilde{v}$. The narrow spectra (bandwidth $\delta\omega$) associated with the modes do not overlap; compare to the spectral support for a conventional pulsed multimode field in Fig.~\ref{Fig:MultimodeWaveguide}(c). (b) Spatio-temporal intensity profiles for the same first four individual modes in Fig.~\ref{Fig:MultimodeWaveguide}(d), but each with bandwidth 0.2~nm (18-ps pulse widths). (c) Spatio-temporal profile of a superluminal ST supermode ($\theta\!=\!75^{\circ}$) formed of the first 8 even-symmetry modes with equal weights. The intensity profile is shown at three axial planes. The X-shaped central peak has a temporal width of 280~fs ($\Delta\lambda\!\approx\!22$~nm at $\lambda_{\mathrm{o}}\!=\!1.55$~$\upmu$m. (d) The time-averaged intensity $I(x,z)$ is invariant along the propagation direction. The axial planes selected in (c) are identified by dashed arrows.}
\label{Fig:BasebandSTSupermode}
\end{figure*}

As mentioned above, the ideal limit for any ST supermode -- whereupon each contributing mode is associated with a single frequency $\omega_{m}$ -- cannot be attained in practice because it requires infinite energy for its realization. Instead, any realizable ST supermode must feature a `spectral uncertainty': a finite bandwidth $\delta\omega$ associated with each mode centered at the ideal frequencies $\omega_{m}$. However, unlike the conventional pulsed multimode field, the bandwidth assigned to each mode is small $\delta\omega\!\ll\!\Delta\omega$ [compare Fig.~\ref{Fig:BasebandSTSupermode}(a) with Fig.~\ref{Fig:MultimodeWaveguide}(c)]. To construct a finite-energy ST supermode, we start -- just as for an ideal ST supermode -- by determining the discrete modal frequencies $\omega_{m}$ at the intersection of the modal dispersion curves $\beta_{m}(\omega)$ with the baseband spectral plane $\omega\!=\!\omega_{\mathrm{o}}+(\beta-nk_{\mathrm{o}})\widetilde{v}$, and then associate a finite bandwidth $\delta\omega$ with each mode (assumed for simplicity to be independent of $m$), rather than taking purely monochromatic modes. Therefore, the spectrum is \textit{not} discrete, and the modal contributions are $E_{m}(x,z,t)\!=\!e^{i\{\beta_{m}(\omega_{m})z-\omega_{m}t\}}u_{m}(x)\psi_{m}(z,t)$, where $\psi_{m}(z,t)$ is the axial modal envelope. As such, the envelope for the finite-energy ST supermode appears at first sight to be similar to the conventional pulsed multimode field in Eq.~\ref{Eq:ModeConventionalPulsed}. However, there are several crucial differences between the two guided pulsed multimode fields. First, whereas the carrier frequency is $\omega_{\mathrm{o}}$ for the pulsed modes in a conventional field, it is $\omega_{m}$ for each mode in the ST supermode. Second, the integration in Eq.~\ref{Eq:ModeConventionalPulsed} for the conventional field is carried out over the full bandwidth $\Delta\omega$ (centered at $\omega_{\mathrm{o}}$), whereas for the ST supermode it extends over only $\delta\omega$ (centered at $\omega_{m}$); see Fig.~\ref{Fig:BasebandSTSupermode}(a). Consequently, the temporal extent of $\psi_{m}(z;t)$ for the ST supermode is $\Delta T_{\mathrm{p}}\!\sim\!\tfrac{1}{\delta\omega}\!\gg\!\Delta T$ [Fig.~\ref{Fig:BasebandSTSupermode}(b)]. Third, the Taylor expansion for $\beta_{m}(\omega)$ for each mode is centered at a different frequency $\omega_{m}$ rather than the same frequency $\omega_{\mathrm{o}}$. However, in most cases the change in values of $\widetilde{v}_{m}$ and $\eta_{m}$ caused by this frequency shift are negligible.

Of course, the attributes of finite-energy ST supermodes also deviate from their ideal counterparts. First, realistic ST supermodes are \textit{not} strictly propagation invariant. At $z\!=\!0$ the profile still has the central X-shaped feature at $t\!=\!0$; however, the profile no longer extends indefinitely along $t$, and is instead limited to a temporal width $\Delta T_{\mathrm{p}}\!\sim\!\tfrac{1}{\delta\omega}$ determined by the axial envelope. Indeed, at $z\!=\!0$ the envelope for the $m^{\mathrm{th}}$ mode is $\psi_{m}(0;t)\!=\!\int_{\delta\omega}\!d\Omega\widetilde{\psi}(\Omega)e^{-i\Omega t}$, which is independent of $m$. Therefore, all the modal envelopes initially overlap, but subsequently walk off with propagation along $z$. Each envelope propagates at a group velocity $\widetilde{v}_{m}$ so that modal GVD broadens the superposed envelope, and causes walk-off with respect to the central X-shaped feature of the ST supermode that travels instead at a group velocity $\widetilde{v}\!=\!c\tan{\theta}$. Moreover, each envelope $\psi_{m}(z;t)$ undergoes waveguide GVD (and potentially chromatic GVD) with a chirp parameter $a_{m}\!=\!\eta_{m}(\delta\omega)^{2}z$, which further contributes to deforming the finite-energy ST supermode with propagation [Fig.~\ref{Fig:BasebandSTSupermode}(c)].

\begin{figure*}[t!]
\centering
\includegraphics[width=18cm]{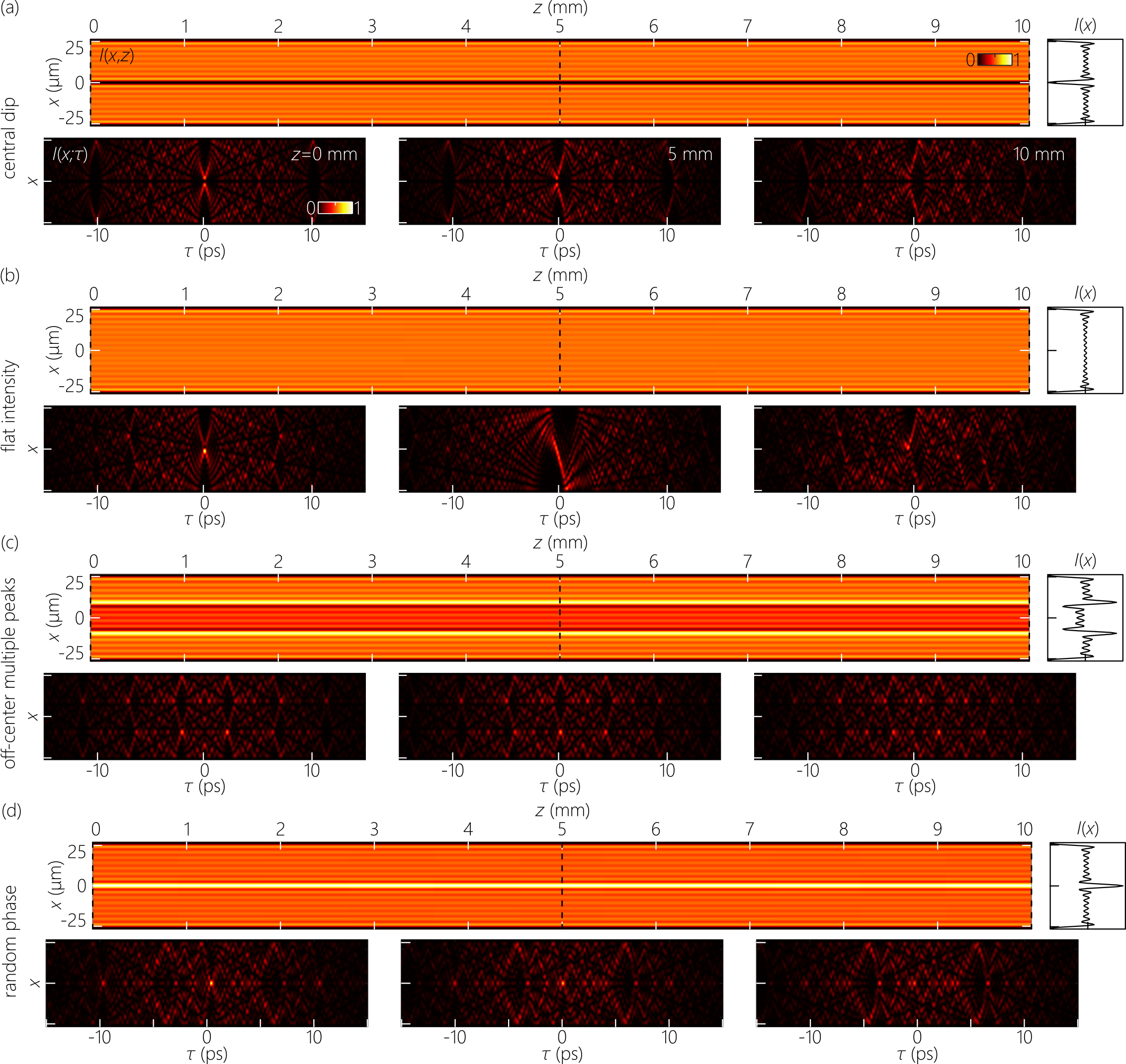}
\caption{Changing the intensity profile of ST supermodes by modal-weight engineering. The value of the parameters for the waveguide and the ST supermodes are the same as in Fig.~ \ref{Fig:BasebandSTSupermode}. (a) Superposing the first 8 odd-parity modes with equal weights results in a null at the center of the intensity profile. (b) Combining the first 16 modes of both odd and even modes creates a uniform field across the waveguide. (c) Superposing modes indexed by $3m+2$ creates two symmetric off center peaks. (d) Combining the first 8 even-parity modes with equal weights but random relative phases selected independently over $[0,2\pi]$. Although the central X-shaped feature is no longer visible in the spatio-temporal profile $I(x;\tau)$ of the ST supermode at any $z$, the time-averaged intensity is nevertheless the same as that in Fig.~\ref{Fig:BasebandSTSupermode}(d).}
\label{Fig:STModalProfiles}
\end{figure*}

Nevertheless, the chirp parameter is reduced with respect to a conventional pulsed multimode field by a factor $(\tfrac{\Delta\omega}{\delta\omega})^{2}$, which can be quite significant. For example, in the proof-of-principle experiment in \cite{Shiri2022STsupermodes} the values were $\Delta\lambda\!\approx\!1.7$~nm and $\delta\lambda\!\approx\!42$~pm, so that GVD is reduced by a factor of $\approx\!1600$. Therefore, both waveguide and chromatic GVD can be drastically reduced in an ST supermode with respect to a conventional pulsed multimode field having the same total bandwidth $\Delta\omega$. However, the modal GVD experienced by the finite-energy ST supermode and the conventional field are the same if the same set of modes contribute to them. Therefore, in practice, ST supermodes can help reduce dispersive effects in those scenarios dominated by chromatic or waveguide GVD, but not those in which modal GVD is the dominant factor.

Although the spatio-temporal intensity profile for finite-energy ST supermodes are \textit{not} propagation invariant [Fig.~\ref{Fig:BasebandSTSupermode}(c)], their time-averaged intensity $I(x,z)\!=\!\int\!dt\,|E(x,z;t)|^{2}$ can nevertheless be axially invariant [Fig.~\ref{Fig:BasebandSTSupermode}(d)]. It is straightforward to show that axial invariance of $I(x,z)$ for a finite-energy ST supermode is guaranteed as long as the spectral domains associated with the contributing modes do not overlap, which is easily achieved when $\delta\omega\!\ll\Delta\omega$. In this case, the time-averaged intensity $I(x,z)\!=\!\sum_{m}|A_{m}|^{2}|u_{m}(x)|^{2}$ retains the same form as in an ideal ST supermode. Axial variation in $I(x,z)$ occurs only when the modal bandwidths overlap spectrally.

\begin{figure*}[t!]
\centering
\includegraphics[width=17.5cm]{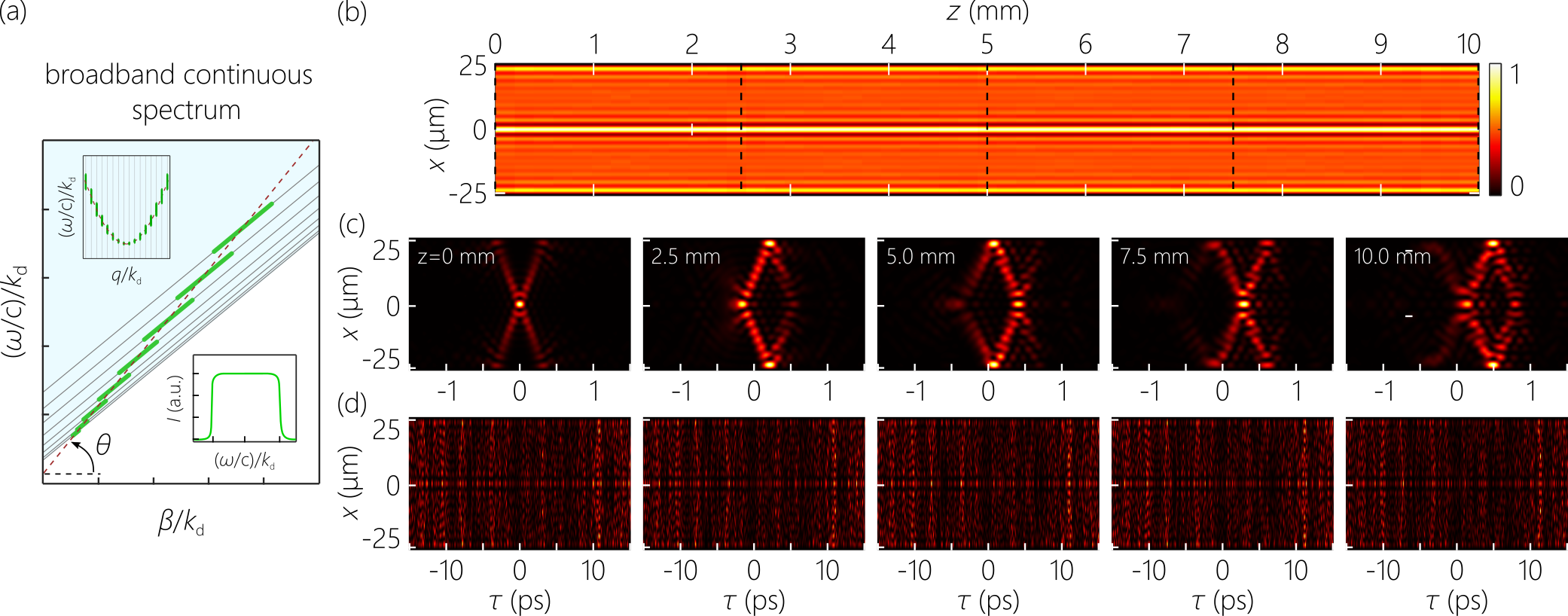}
\caption{Spectrally continuous ST supermodes. (a) ST supermode spectrum projected onto the $(\beta,\tfrac{\omega}{c})$-plane. The \textit{full} input spectrum of the pulse coupled to the waveguide is accommodated by the waveguide modes selected. Instead of assigning a constant narrow bandwidth to each mode that is smaller than the mode spacing, the assigned spectra are contiguous between the successive modes. The top-left inset shows the spectrum projected onto the $(q,\tfrac{\omega}{c})$-plane, and the bottom-right inset shows the overall spectral magnitude ($k_{d}\!=\!\tfrac{\pi}{d}$). (b) The time-averaged intensity profile $I(x,z)$ with $\theta\!=\!75^{\circ}$ and selecting the first 8 even-parity modes. (c) $I(x;\tau)$ at different axial planes; only the central X-shaped feature initially appears, which undergoes deformation along $z$ because of modal dispersion. (d) Applying a random phase to each frequency spreads the field out temporally and randomises the intensity distribution. Nevertheless, the time-averaged intensity remains that in (b).}
\label{Fig:broadbandincoherent}
\end{figure*}

\subsection{Sculpting the transverse spatial profile}

Equation~\ref{Eq:TimeAveragedIntensityAxiallyInv}, which applies to both ideal and finite-energy ST supermodes, indicates that the propagation-invariant time-averaged spatial intensity profile in the waveguide can be tuned by varying the modal weights in the ST supermode. In Fig.~\ref{Fig:BasebandSTSupermode}(d), we made use of equal-weighted modes with even-parity symmetry ($m\!=\!0,2,4,\cdots$), which results in an on-axis peak atop an approximately flat background. Alternatively, one may produce a propagation-invariant on-axis \textit{dip} by selecting modes with odd-parity symmetry ($m\!=\!1,3,5,\cdots$) as shown in Fig.~\ref{Fig:STModalProfiles}(a). The evolution of the spatio-temporal profile resembles that in Fig.~\ref{Fig:BasebandSTSupermode}(d) except that an intensity null is always maintained at $x\!=\!0$, thereby resulting in an on-axis dip in the time-averaged intensity at $I(x\!=\!0,z)$. This unique dark-beam profile in a multimode waveguide offers opportunities for atom guidance in hollow-core waveguides and in reducing unwanted nonlinear interactions in the transport of high-energy pulses.

Alternatively, by selecting equal-weighted modes with both even- and odd-parity symmetry ($m\!=\!0,1,2,3\cdots$), we obtain a flat intensity profile across the waveguide cross section, as shown in Fig.~\ref{Fig:STModalProfiles}(b). Such a configuration can perhaps be useful in lighting applications (see below for a discussion of spectrally incoherent fields). A further example is shown in Fig.~\ref{Fig:STModalProfiles}(c) in which a symmetric double-peaked structure is realized by selecting a subset of even- and odd-parity modes indexed by $3m+2$. Using Fourier analysis, more complex intensity profiles can be synthesized by carefully controlling the modal weights $|A_{m}|^{2}$. In general, however, only even-parity intensity distributions can be produced in a symmetric waveguide using this procedure.

\subsection{Impact of spectral coherence}

Examining Eq.~\ref{Eq:TimeAveragedIntensityAxiallyInv} shows that the relative modal phases do \textit{not} impact the intensity profile $I(x,z)$. This indicates that random phases introduced between the modal weights (which can arise when using a spectrally incoherent source rather than a coherent pulse) will not induce axial changes as long as the relative magnitudes are maintained constant. We confirm this conclusion in Fig.~\ref{Fig:STModalProfiles}(d) where we introduce random phases selected with a uniform probability from the span $[0,2\pi]$ to the modes comprising the ST supermode. The intensity $I(x,z)$ is identical to the coherent field in Fig.~\ref{Fig:BasebandSTSupermode}(d), where the same constitutive modes were in phase. However, the spatio-temporal profile $I(x,z;t)$ is of course impacted by these random phases. This result indicates that even spectrally incoherent light can be used to produce ST supermodes.

\section{Spectrally continuous ST supermodes}

Neither ideal ST supermodes (where the spectrum is discretized at frequencies $\omega_{m}$) nor finite-energy ST supermodes (where narrow spectra $\delta\omega$ are centered at $\omega_{m}$) utilize the full spectrum $\Delta\omega$. Therefore, this overall approach to synthesizing ST supermodes is not spectrally nor energy efficient. Surprisingly, one may indeed construct \textit{spectrally continuous} ST supermodes while retaining some of the advantages of their ideal counterparts. The general approach is illustrated in Fig.~\ref{Fig:broadbandincoherent}(a), where the full spectrum $\Delta\omega$ is divided amongst $M$ spectrally contiguous modes, each of bandwidth $\Delta\omega_{m}\!\sim\!\tfrac{\Delta\omega}{M}$. As long as the modal spectra do not overlap, the time-averaged intensity remains $I(x,z)\!=\!\sum_{m}|A_{m}|^{2}|u_{m}(x)|^{2}\!=\!I(x,0)$, which is axially invariant as shown in Fig.~\ref{Fig:broadbandincoherent}(b).

Of course, the time-resolved spatio-temporal profile $I(x,z;t)$  is \textit{not} invariant along $z$. Interestingly, as a consequence of the absence of gaps in its spectrum, the temporal extent of the ST supermode is now reduced to encompass only the central X-shaped feature, and thus has a similar profile to its freely propagating counterpart [Fig.~\ref{Fig:broadbandincoherent}(c)]. Furthermore, this finite-energy ST supermode undergoes dispersion with propagation, with waveguide and chromatic GVD reduced by a factor $\sim\!M^{2}$, which can be significant. Just as in the case of finite-energy ST supermodes examined above, modal GVD is the same as that for a conventional pulsed multimode field comprising the same modes. Therefore, spectrally continuous ST supermodes provide an advantage with regards to dispersion reduction without sacrificing power, in addition to the axial invariance of the time-averaged intensity.

The axial invariance of $I(x,z)$ extends to spectrally incoherent light as well, as shown in Fig.~\ref{Fig:broadbandincoherent}(d). Here we assign independent randomly selected phases with a uniform probability over $[0,2\pi]$, with each phase assigned to a spectral window of width $\delta\lambda\!=\!30$~pm. Consequently, the field spreads temporally with no discernible spatio-temporal features as expected for an incoherent field, in contrast to the coherent scenario in Fig.~\ref{Fig:broadbandincoherent}(c). Nevertheless, the time-averaged intensity $I(x,z)$, which is here the quantity of interest, remains axially invariant as shown in Fig.~\ref{Fig:broadbandincoherent}(b). This class of spectrally continuous ST supermodes has not yet been demonstrated experimentally.

\section{Discussion}

Although ST supermodes can be considered guided counterparts of freely propagating STWPs, there are nevertheless significant differences between them. First, whereas the spectrum of ideal ST supermodes is discrete, that for an ideal STWP is continuous. Second, the maximum propagation distance $L_{\mathrm{max}}$ of realistic STWPs in free space is limited by the spectral uncertainty, whereas that for ST supermodes is determined by the spectral overlap between the underlying modes. Therefore, even in presence of spectral uncertainty in a realistic ST supermode (the finite spectrum associated with each mode), its transverse intensity profile can remain invariant indefinitely. Third, the spectral discretization results in an extended spatio-temporal structure for ST supermodes around a central X-shaped feature, whereas freely propagating STWPs comprise only an X-shaped feature. The impact of spectral discretization on broadening the spatio-temporal profile occurs even in free STWPs as confirmed recently in the context of ST Talbot effects \cite{Yessenov2020PRLveiled,Hall2021APLSTTalbot}. On the other hand, we have shown here that taking a spectrally continuous ST supermode has the same spatio-temporal profile as its free STWP counterpart. Finally, whereas the spatio-temporal profile of a free STWP remains invariant until $L_{\mathrm{max}}$ is reached, the corresponding profile for an ST supermode is deformed because of modal dispersion even though the time-averaged intensity profile is invariant. Where the impact of waveguide and chromatic GVD can be significantly reduced in a realistic ST supermode, the impact of modal dispersion remains.

The theoretical treatment provided here extends beyond the constraints we have imposed on our analysis. First, although we focused here on coherent pulsed light, the main conclusions remain valid for broadband, spectrally incoherent continuous-wave radiation. The same conclusion was reached for free STWPs (see \cite{Turunen2008OE,Saastamoinen2009PRA,Turunen2010PO} for theoretical treatments and \cite{Yessenov2019Optica,Yessenov2019OL} for experimental realizations). This opens up new opportunities for delivering broadband incoherent light with controllable intensity profiles over multimode fibers for lighting applications. Second, with regards to polarization, all our calculations here pertain to TM-polarized modes, but the analysis applies equally to TE and mixed polarization modes. Third, we have considered here index-guided ST supermodes. Nevertheless, this basic concept applies to any other guidance mechanism; e.g., we outline in the Appendix the theory of ST supermodes in planar-mirror waveguides, and \cite{Guo2021PRR} describes the scenario of a graded-index planar waveguide. Finally, our analysis assumed planar waveguides with light confined along one transverse dimension. However, the analysis may be extended to encompass ST supermodes in conventional waveguides in which light is confined in both transverse dimensions, and also to fiber-guided ST supermodes. Recent theoretical \cite{Guo2021Light,Pang2021OL} and experimental \cite{Yessenov2022NatComm,Pang22OL,Yessenov22OL} work on the synthesis of STWPs localized in all dimensions now brings this possibility closer to experimental realization.

A recent breakthrough \cite{Stefanska22arxiv} was achieved with respect to the nonlinear synthesis of ST supermodes in a multimode optical fiber via a high-energy excitation pulse \cite{Kibler2021PRL}. However, because the spectrum associated with each mode in the ST supermode is large, the defining characteristics of ST supermodes as emerging from our analysis here are not expected to be observable. Finally, we note that the structure of ST supermodes is similar to that of the field in a waveguide moving at relativistic speeds \cite{Qu2016JOSAB}. This can be understood by recognizing that STWPs can be obtained from the Lorentz transformation of monochromatic beams \cite{Belanger1986JOSAA,Longhi2004OE,Saari2004PRE,Yessenov22Lorentz}. The study of STWPs and ST supermodes can thus enrich our understanding of relativistic interactions between optical fields and photonic devices.

\section{Conclusions}

In conclusion, we have presented a theoretical treatment for ST supermodes in planar multimode waveguides, which are the guided counterparts to freely propagating STWPs. In contrast to conventional pulsed multimode fields in waveguides or fibers, ST supermodes retain their time-averaged intensity profile indefinitely. Moreover, the group index of the ST supermode can be tuned above or below the group indices of its constitutive modes. It is expected that the theoretical treatment developed here and recently reported experimental results \cite{Shiri2022STsupermodes} can be extended to realizing ST supermodes in multimode fibers or conventional waveguides in which the field is confined in both transverse dimensions.


\begin{figure*}[t!]
\centering
\includegraphics[width=18cm]{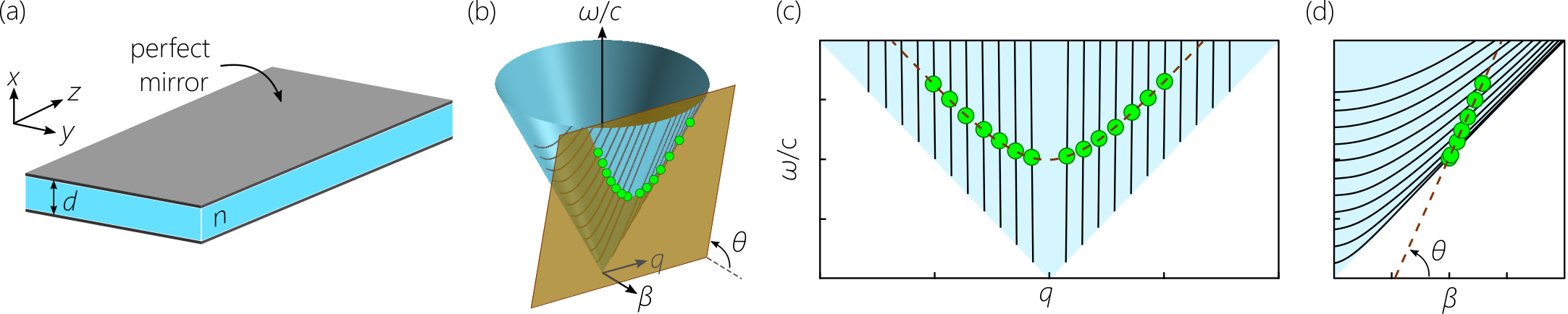}
\caption{\textbf{Planar-mirror waveguide.} (a) Schematic of planar-mirror waveguide structure. (b) The modal dispersion curves are 1D curves on the surface of the light-cone $q^{2}+\beta^{2}\!=\!(n\tfrac{\omega}{c})^{2}$, where $q$ and $\beta$ are the transverse and axial wave numbers, respectively. (c) The spectral support for a baseband ST supermode is the intersection of a tilted plane $\mathcal{P}(\theta)$ with the modal dispersion curves, which corresponds to a collection of points (shown by green dots) in the $(q,\tfrac{\omega}{c})$-plane and (d) in the $(\beta,\tfrac{\omega}{c})$-plane. Here $\theta\!=\!60^{\circ}$ and $n\!=\!1.5$, which corresponds to a superluminal ST supermode.}
\label{Fig2:mirror_waveguide}
\end{figure*}

\vspace{2mm}
\noindent\textbf{Funding.} U.S. Office of Naval Research (ONR) N00014-17-1-2458 and N00014-20-1-2789.




\section*{Appendix: ST supermodes in planar-mirror waveguides}

\begin{figure*}[t!]
\centering
\includegraphics[width=18cm]{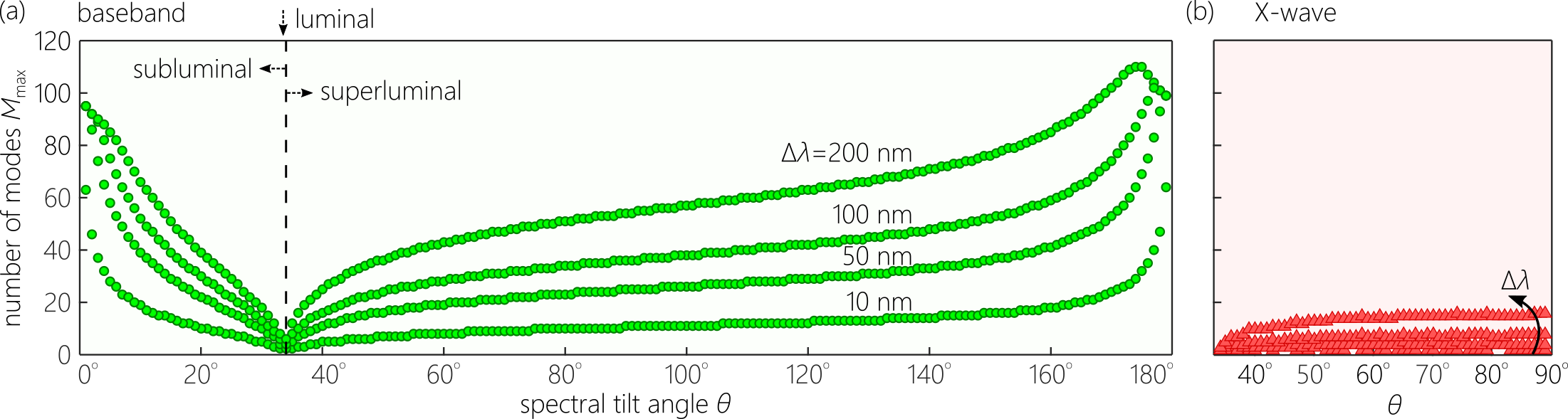}
\caption{Maximum number of modes accessible in a planar-mirror waveguide with $\theta$ for (a) baseband ST supermodes at bandwidths $\Delta\lambda\!=\!10$, 50, 100, and 200~nm; $\lambda_{\mathrm{o}}\!=\!1.55$~$\upmu$m, $n\!=\!1.5$, and $d\!=\!50$~$\upmu$m. (b) Same as (a), but for X-wave ST supermodes. For X-waves, $\theta$ is restricted to the superluminal region and $\theta\!=\!90^{\circ}$.}
\label{Fig:STSupermodeNumberOfModes_mirror}
\end{figure*}

For sake of comparison to the index-guided ST supermodes described in the main text, we consider their counterpart in a planar waveguide formed of two perfect parallel mirrors [Fig.~\ref{Fig2:mirror_waveguide}(a)]. The separation between the two mirrors along $x$ is $d$, and the refractive index of the medium is $n$ (assumed to be independent of frequency). The quantized transverse wave number $q_{m}$ for the $m^{\mathrm{th}}$ mode ($m\!=\!1,2,3,\cdots$) is $q_{m}\!=\!m\tfrac{\pi}{d}\!=\!k\sin{\varphi_{m}}\!=\!mk_{\mathrm{c}}$, where $k\!=\!nk_o=\tfrac{n\omega}{c}$, $\omega_{\mathrm{c}}\!=\!\tfrac{\pi c}{nd}$ is the cut-off frequency, $\lambda_{\mathrm{c}}\!=\!2d$ is the cut-off wavelength in the waveguide, $k_{\mathrm{c}}\!=\!\tfrac{2\pi}{\lambda_{\mathrm{c}}}\!=\!n\tfrac{\omega_{\mathrm{c}}}{c}$, $\varphi_{m}$ is the frequency-dependent bounce angle of a plane wave relative to the $z$-axis \cite{Saleh2007Book}, and the axial wave number is $\beta_{m}\!=\!\sqrt{n^{2}k_{\mathrm{o}}^{2}-q_{m}^{2}}\!=\!k\cos{\varphi_{m}}$ where $k_{\mathrm{o}}\!=\!\tfrac{2\pi}{\lambda_{\mathrm{o}}}$ and $\lambda_{\mathrm{o}}$ is the vacuum wavelength. The mode profiles are $u_{m}(x)\!\propto\!\cos{(q_{m}x)}$ for odd $m$ and $u_{m}(x)\!\propto\!\sin({q_{m}x)}$ for even $m$, are independent of $\omega$, and orthonormal $\int\!dx\,u_{m}(x)u_{n}^{*}(x)\!=\!\delta_{nm}$. The group velocity at $\omega\!=\!\omega_{\mathrm{o}}$ is $\widetilde{v}_{m}\!=\!1\big/\tfrac{d\beta_{m}}{d\omega}\big|_{\omega_{\mathrm{o}}}\!=\!\tfrac{c}{n}\cos{\{\varphi_{m}(\omega_{\mathrm{o}})\}}$, which is always subluminal. The associated GVD coefficient is $\eta_{m}\!=\!\tfrac{d^{2}\beta_{m}}{d\omega^{2}}\big|_{\omega_{\mathrm{o}}}\!=\!-\tfrac{n^2m^2\pi^2}{c^2d^2\beta^3_m}\!=\!-\tfrac{n}{c\omega_{\mathrm{o}}}\tan^{2}{\varphi_{m}}\sec{\varphi_{m}}$; the GVD of each mode is always anomalous (negative-valued $\eta_{m}$), and increases in absolute value with modal index; $\varphi_{m}\!\rightarrow\!90^{\circ}$ for higher $m$, whereby $|\eta_{m}|$ increases. In general, the temporally resolved and time-averaged intensity profiles have similar forms to those for index-guided ST supermodes.

\subsection{X-wave ST supermodes}

The frequencies $\omega_{m}$ for an X-wave ST supermode in the planar-mirror waveguide are at the intersection of the dispersion curves $\beta_{m}(\omega)$ with the plane $\omega\!=\!\beta\widetilde{v}$, where $\widetilde{v}\!=\!c\tan{\theta}$, $\theta$ is the spectral tilt angle, and we define a group index $\widetilde{n}\!=\!c/\widetilde{v}\!=\!\cot{\theta}\!<\!n$. The intersection points are at $\omega_{m}\!=\!m\omega_{\mathrm{c}}\tfrac{n}{\sqrt{n^{2}-\widetilde{n}^{2}}}$ and $\beta_{m}\!=\!mk_{\mathrm{c}}\tfrac{\widetilde{n}}{\sqrt{n^{2}-\widetilde{n}^{2}}}$. The group velocity for any mode $\widetilde{v}_{m}\!=\!\tfrac{c}{n}\sqrt{1-(\tfrac{q_{m}}{n\omega_{\mathrm{c}}/c})^{2}}\!=\!\widetilde{v}_{\mathrm{g}}$ is independent of $m$, is subluminal, and $\widetilde{v}\cdot\widetilde{v}_{\mathrm{g}}\!=\!(\tfrac{c}{n})^{2}$; whereas the ST supermode is superluminal at $\widetilde{v}$. Nevertheless, the GVD parameters for the modes $\eta_{m}$ remain different. Moreover, the phase velocity of the modes $v_{\mathrm{ph}}\!=\!\tfrac{\omega_{m}}{\beta_{m}}\!=\!\widetilde{v}$ is equal for all the modes, and is equal to the group velocity for the ST supermode. 

Because the frequencies of waveguide modes $\omega_{m}$ contributing to this ST supermode are harmonics of the fundamental frequency $\omega_{\mathrm{c}}$ ($\omega_{m}\!\propto\!m\omega_{\mathrm{c}}$), a large bandwidth is required to produce an ST supermode with a reasonable number of underlying modes (see \cite{ZamboniRached2001PRE,ZamboniRached2002PRE,ZamboniRached2003PRE}). This makes the synthesis of ST supermodes based on X-waves difficult, just as in their index-guiding counterparts. 

\subsection{Baseband ST supermodes}

For baseband ST supermodes the modal frequencies $\omega_{m}$ are at the intersection of the dispersion curves $\beta_{m}(\omega)$ with the plane $\omega\!=\!\omega_{\mathrm{o}}+(\beta-nk_{\mathrm{o}})c\tan{\theta}$,
\begin{equation}
  \frac{n\omega_m}{c}= \frac{k}{(\widetilde{n}-n)} \left\{\widetilde{n} \pm n\sqrt{1+\left(\frac{m \lambda_o}{2 n d}\right)^2 \frac{n-\widetilde{n}}{n+\widetilde{n}} }\right\}, 
  \label{eq:mode_freq_baseband}
\end{equation}
with $\widetilde{v}\!=\!c\tan{\theta}$ and $\widetilde{n}\!=\!\cot{\theta}$, which can be superluminal, subluminal or even negative-valued. There is no limit on $m$ in the superluminal regime ($\widetilde{n}\!<\!n$), in the subluminal regime ($\widetilde{n}\!>\!n$) real solutions for $\omega_{m}$ require that $m<\tfrac{2nd}{\lambda_{\mathrm{o}}}\sqrt{\frac{|n-\widetilde{n}|}{n+\widetilde{n}}}$.

\subsection{Sideband ST supermodes}    

We now consider sideband ST wave packets whose modal centers lie in the ($\beta,\tfrac{\omega}{c})$-plane along the line $\tfrac{\omega_{m}}{c}\!=\!k_{\mathrm{o}}+(\beta_{m}(\omega_{m})+nk_{\mathrm{o}})\tan{\theta}$. The group velocity of the ST supermode is $\widetilde{v}\!=\!c\tan{\theta}$. Similar to ST supermodes based on X-waves, those based on sideband STWPs are also only superluminal. 

\begin{equation}
\frac{n\omega_m}{c}= \frac{k}{(\widetilde{n}+n)} \left\{\widetilde{n}\pm n\sqrt{1+\left(\frac{m\lambda_{\mathrm{o}}}{2nd}\right)^2\frac{n+\widetilde{n}}{n-\widetilde{n}}}\right\}. 
\label{eq:mode_freq_sideband}
\end{equation}
In general, like the X-wave, a sideband ST supermode requires a large bandwidth that is not practical in optics.  

\subsection{Comparison of X-wave, baseband, and sideband ST supermodes}
To conclude this review of planar-mirror waveguide supermode properties we compare the number of modes that can be accessed for baseband and X-wave ST wave packets. Mode frequency equations for X-waves and baseband waves (given in Appendix A and B)  can be used to find the number of modes, $\Delta m$, that are present for a given fixed bandwidth $\Delta\omega$. Fig. \ref{Fig:STSupermodeNumberOfModes_mirror} shows calculations of the number of modes available for supermode propagation at a given spectral tilt angle and bandwidth. In all cases baseband supermodes have many more modes available than X-wave supermodes. We also note that baseband modes have low $m$ values while X-waves must use high $m$ value modes (i.e., they are highly nonparaxial).

\bibliography{diffraction.bib}
\end{document}